# Virtual On-demand Test Lab using Cloud based Architecture

Rohit Kewlani
Adobe Systems
I-1A, City Center, Sector 25-A
Noida, U.P.
India
rkewlani@adobe.com

## ABSTRACT

*Over a past few decades, VM's or Virtual machines have sort of gained a lot of momentum, especially for large scale enterprises where the need for resource optimization & power save is humongous, without compromising with performance or quality. They are a perfect environment to experiment with new applications/technologies in a completely secure and closed environment. This paper discusses how the VM technology can be leveraged to solve day to day requirement of an odd hundreds or thousands of people, organization-wide, with new computational resources using a cluster of heterogeneous low or high-end machines, independent of underlying OS, thereby maximizing resource utilization. It takes into account both opensource (like VirtualBox) & other proprietary technologies (like VMWare Workstations) available till date to propose a viable solution using cloud computing concept. The ease of scalability to multiple folds for optimizing performance & catering to an even larger set are some of the salient features of this approach. Using the snapshot feature, the state of any VM instance could be saved & served back again on request. Now, this implementation is also served by VMWare ESX server but again it's a costly solution & requires dedicated high-end machines to work with.*

## 1. INTRODUCTION

The need for a powerful & flexible execution environment has been topping the priority list for most of the organizations in existence till date. Modern day CPU's, that are used in general day to day computing are very much capable enough to serve more than just one user at a time. With the advent of VM technology, harnessing the true power of underlying hardware isn't a farfetched dream any more. We already have opensource technologies like VirtualBox & VMware Workstation that are extensively used every day but none have been thought of as a building block to something like big VM pool. ViTL or Virtual Testing Lab is a framework having a repository of VM's & is based on a service-oriented architecture, wherein requests can be made from a web-based interface. As soon as a request is made, the framework based on a customizable load-balancing algorithm, decides which host machine on the cloud would handle the request. Once the host has been chosen, the clone of the requested VM is instantiated on it. Finally, when the VM is up & running, details such as I/P address, a default user/password to connect is shared with the requestor. Note that this approach can be taken one step ahead to handle automated requests, to run a bunch of tasks on a targeted VM instance. To start off with, we'll discuss in detail exactly how this complete process works, starting from request submission till the part where VM details being shared with the user.

## 2. How does this work ?

The complete workflow has been broken down into 4 sub-sections. In the following section 2.1 we'll understand how exactly the web requests are handled

### 2.1 Web-request handling

The very first step is to show the end-user an interface through which the request can be submitted. Here, a list of supported distributions (linux/windows/mac variants) can be shown on a web interface by querying a database at the backend. In the figure below is a page created in PHP which does exactly the same.

*Figure 2.1.1 Web Request Page*

Herein a set of inputs can be taken from the end user like which distribution is being requested for. Next thing could be the underlying processor i.e. INTEL/AMD etcetera. Note that both VMWare & VirtualBox emulate the OS instance based on the underlying hardware on top of which they are running. A username field identifies the requestor, which again can be populated through a request to an authentication server (e.g. LDAP authentication). The lease time option allots the VM for a stipulated time frame after which it shuts down the instance & resources are freed.

On submitting the request these fields are automatically stored in a database to generate a unique request ID, which would be used for any future references.

```
+-------------+---------------------------------+
| Field       | Type                            |
+-------------+---------------------------------+
| VmID        | int(10)                         |
| OSName      | varchar(30)                     |
| CloneVMXPath| varchar(100)                    |
| OSFamily    | enum('WIN','LINUX','MAC','OPEN SOLARIS') |
| DisplayName | varchar(100)                    |
+-------------+---------------------------------+
```
*Figure 2.1.2 VM repository database*

### 2.2 Setting up the run configuration

This section will briefly go through the setup phase.

Next thing in queue is to decide upon which host to choose to serve the user request. Here comes into picture the concept of load-balancing. Below is the table which describes the configuration of various hosts in the cloud.

```
+----------------+---------------------+
| Field          | Type                |
+----------------+---------------------+
| node_id        | int(10) unsigned    |
| ip_addr        | varchar(20)         |
| hostname       | varchar(64)         |
| distro_name    | varchar(128)        |
| architecture   | varchar(8)          |
| mac_addr       | varchar(20)         |
| total_mem      | int(11)             |
| avail_mem      | int(11)             |
| machine_status | varchar(20)         |
| timestamp      | timestamp           |
| cpu_model      | enum('INTEL','AMD') |
| runningvms     | int(10) unsigned    |
| automation     | enum('Y','N','B')   |
+----------------+---------------------+
```
*Figure 2.2.1 Host configuration database*

Every host that is a part of the cloud registers itself with the following configurational details :

- I/P address for communicating with the host.
- Hostname

- Distribution name (win/mac/linux)
- Mac or Ethernet address (48-bit)
- Total Memory or RAM
- Current freely available memory
- Status of the host i.e. offline/online (shall be discussed later in detail)
- TimeStamp when host was last communicated.
- CPU model i.e. INTEL/AMD etcetera.
- Currently running VM instances on the host.
- Whether host would serve user requests for live instances or automation requests or both.

Figure 2.2.2 shows how the set of eligible hosts is decided upon.

There a basically two sets of hosts:
- SET I has hosts which are completely free & have no VM's running currently.
- SET II has a set of hosts which have atleast one VM instance running on them.

If SET I ≠ {empty set} implies we have to choose from SET I. Otherwise, choice has to be made from SET II.

For SET I, the most eligible host is given by formula,

$$\frac{TotalFreeMemory}{TotalMemory}$$

Host whose value for the above expression is maximum is our target host.
The above expression can be further modified to include CPU speed and various other configurational parameters to come up with a more accurate figure.

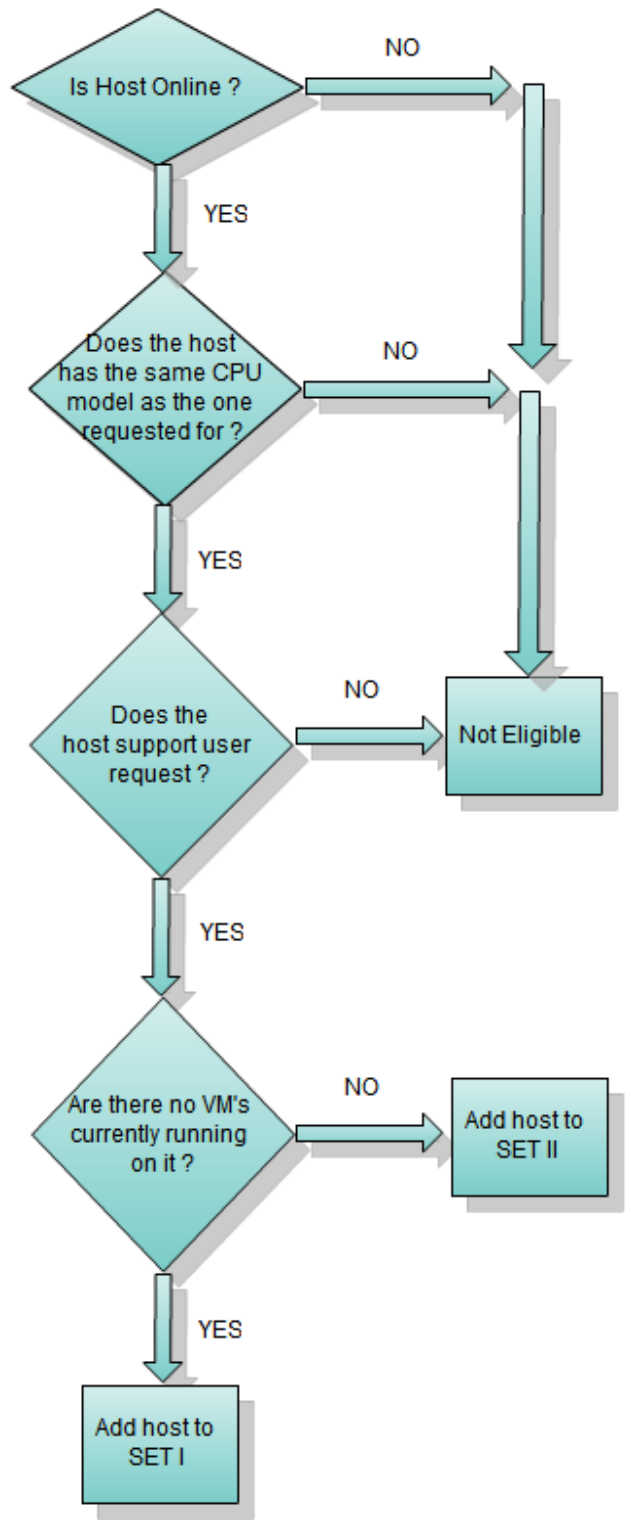

*Figure 2.2.2 Finding set of eligible hosts*

Similarly, for SET II we'll be finding two factors viz. MemoryFactor & VMdistributionFactor.
- MemoryFactor means what percentage of total memory is free on the given host
- VMdistributionFactor means what percentage of total VM's running on the cloud are running on the given host.

Mathematically,

$$MemoryFactor = \frac{Total\ free\ memory}{Total\ Memory}$$

$$VMDistributionFactor = \frac{Total\ VM's\ Running\ on\ the\ host}{Total\ VM's\ Running\ on\ the\ cloud}$$

Finally, the target host is chosen using the formula

$$K * \frac{MemoryFactor}{VMdistributionFactor}$$

In simple terms, the host which has the maximum available free memory & is comparatively the least loaded amongst all. The constant K decides which of the two factors used has more weightage. K=1 should suffice in most cases but can be varied accordingly to optimize distribution. Sometimes, it might be possible that none of the hosts chosen is free or the ones available don't have sufficient enough memory available to trigger one more instance. In such scenarios, the requestor can be notified about the delay in serving & the request can be queued up to be dealt later.

```
+------------------+---------------------------------------------------------------------------------------------------------------------------------------------------------+
| Field            | Type                                                                                                                                                    |
+------------------+---------------------------------------------------------------------------------------------------------------------------------------------------------+
| JobID            | int(10) unsigned                                                                                                                                        |
| AssignedOn       | timestamp                                                                                                                                               |
| Requestor        | varchar(30)                                                                                                                                             |
| IPAddress        | varchar(30)                                                                                                                                             |
| VMXPath          | varchar(100)                                                                                                                                            |
| Status           | enum('NEW','AUTHORIZED','UNAUTHORIZED','PROCESSING','PROCESSED','DELETED','INCOMPLETE','ASSIGNED','UNASSIGNED','LIVE','STOPPED','REMINDER1','REMINDER2' |
| RequestType      | enum('USER','TOD')                                                                                                                                      |
| AuthenticationToken | varchar(30)                                                                                                                                          |
| NodeID           | int(10) unsigned                                                                                                                                        |
| VITLLogFilePath  | varchar(100)                                                                                                                                            |
| VmID             | int(10) unsigned                                                                                                                                        |
| StatusCopyVM     | enum('PASSED','PENDING','FAILED')                                                                                                                       |
| StatusVMUp       | enum('PASSED','PENDING','FAILED')                                                                                                                       |
| StatusIPSet      | enum('PASSED','PENDING','FAILED')                                                                                                                       |
| StatusEmailSent  | enum('PASSED','PENDING','FAILED')                                                                                                                       |
| Architecture     | enum('INTEL','AMD')                                                                                                                                     |
| VMUser           | varchar(20)                                                                                                                                             |
| StatusReason     | varchar(500)                                                                                                                                            |
| LeaseTime        | int(3) unsigned                                                                                                                                         |
| TimeRemaining    | int(20) unsigned                                                                                                                                        |
+------------------+---------------------------------------------------------------------------------------------------------------------------------------------------------+
```

*Figure 2.2.2: Request details database*

### 2.3 Setting up the host

Now that the target host has been chosen, the next step was to configure the host for the VM instance run. We had two options:
- Create a linked clone of the base VM on the fly on the host itself.
- Copy an already created linked clone onto the host.

The first option was easier to implement but creating a clone at runtime was a time-taking process. The other option, to copy the linked clone on request had one major drawback i.e. a linked clone is always linked to it's base VM. Generally the complete path of the base VM is hardcoded into the clone's configuration file. To keep the path consistent across platforms we decided to keep all the base VM's & corresponding linked clones on a common shared file server, capable enough to handle large requests at a time. This shared file server would be mounted on same location on all the hosts, so that the linked clone always points to the same base VM, whatever be the target host. Now at first it seemed it would prove to be a performance bottleneck i.e. keeping everything on a shared file server & launching the clones from there itself, involving network intensive operations everytime. But on the contrary, over a 100 Mbps LAN this difference hardly had any effect. Moreover, managing resources got a lot easier as everything was at one place. Also, dependency on hosts in terms of storage space was no more a matter of concern. For security reasons, permissions can be set on specified hierarchy on the file server i.e. only those files/directories are accessible which are intended to be used for serving request.

All the resources required by the host are copied into a folder (named by a unique id) on the shared file server. The host is then notified about the incoming request & asked to turn on the VM instance, through a handshaking mechanism which shall be discussed shortly.

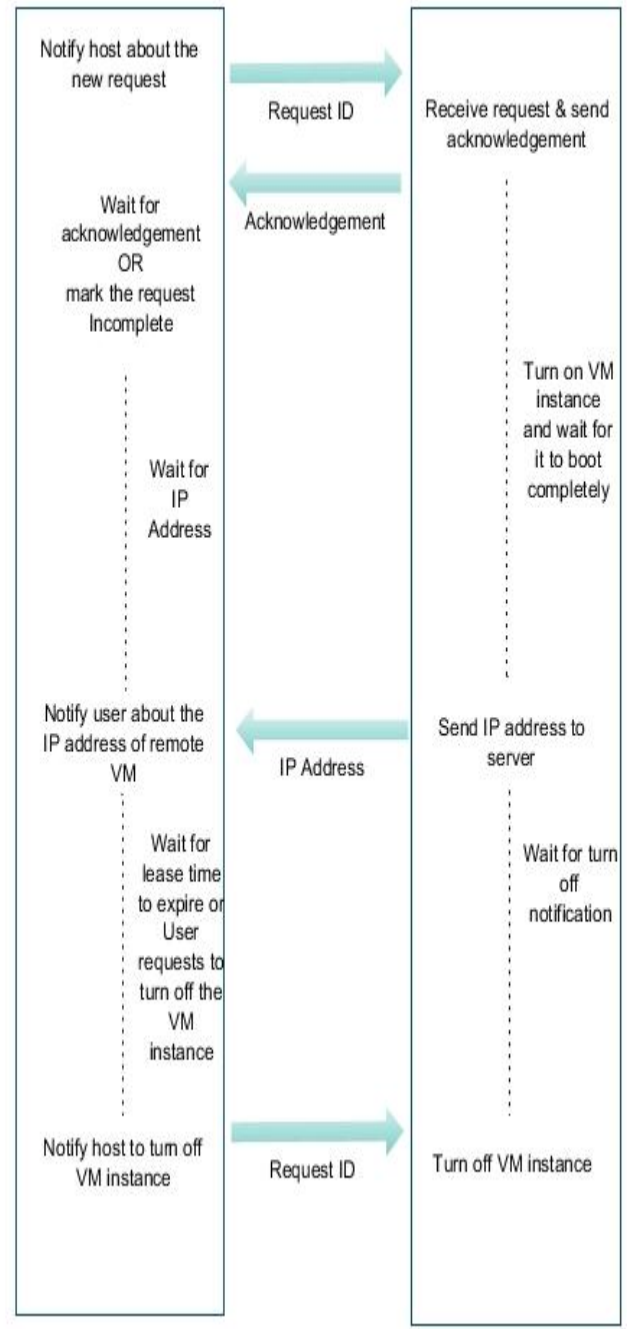

*Figure 2.3.1 Server host handshaking*

### 2.4 Turning on the VM instance

Once a host has been notified, it boots up the requested VM. The IP address can easily be queried through inbuilt utilities like ipconfig (on windows) or ifconfig (on linux/mac) inside the VM.

If the source VM cannot be booted up or if the IP address returned is invalid or the host goes down for some reason, the server is notified & it marks the VM request as incomplete. In such scenarios, a new request is generated with the same configurational details & a new host is assigned using the same procedure as discussed in section 2.2 , except this time the defaulter host won't be a contender anymore.

### 2.5 Sending notification to the requestor

The IP address generated can be shared with the end-user along with a default username/password, which could be used to connect to the remote VM instance. On windows, remote desktop can be done using rdesktop/mstsc. Similarly, for win/mac nxclient can be used, a freeware which can be downloaded from [http://www.nomachine.com/download.php](http://www.nomachine.com/download.php).
Nxclient requires a nxserver to be running on the remote machine. Details regarding configuring nxserver have been shared in section 4.

## 3  Configuring a VM instance

Any new VM which needs to be integrated with the framework must have some pre-configurations done.
- Auto-login set with a default username/password i.e. as soon as the VM instance boots up, it should login automatically.
- If it's a windows/mac VM Nxserver must be installed. Ensure that the nxservice starts successfully on login.
- Firewall should be disabled or configured accordingly to allow remote login.
- Install virtual box or VMware tools inside the VM. Both of them are freely available on the internet.
- Turn off screensaver(if any).
- Disable auto updates to keep disable any updates while usage.

## 4  Adding a new host to the cloud
Any new host that needs to be added must meet a certain set of requirements, before it can be integrated with the system.
- It must have VMware workstation or Virtual Box pre-installed.
- Network connectivity.
- Minimum 512 MB RAM.
- SOAP modules installed to interface with the main server.
- Nxserver for linux/mac

A script can be scheduled on this host which reports back the status of the machine to the main server, maybe a SOAP client using a certain set of exposed SOAP calls.

This scripts reports the following data with every status ping,
- IP address or hostname
- Distribution/Operating System name
- CPU type (Intel/AMD)
- CPU architecture (32-bit/64-bit)
- Total system memory
- Total available free memory

With every ping the host confirms it's presence & sets status as online. If for a time difference, say 5 minutes or more, there's no response from the host, the server marks the host offline.

## 5  Conclusion

We studied the load distribution across a local network with 100Mbps speed, VM instances with maximum memory set to 1GB, five 32-bit Intel based hosts, including one high-end server capable of supporting upto fifteen VM instances at a time & rest four as low-end machines which could run upto 2 instances at a time.

Figure 6.1 shows a block diagram of how the whole system looks like. We can see that initially up till five requests the turnaround time is almost constant. This behaviour can be attributed to load distribution logic, when atleast one host is available with no load. As the load increases, the slope of the graph increases. Finally after fifteen instances have been instantiated, the turnaround time takes further hit as the hosts almost touch their load capacity.

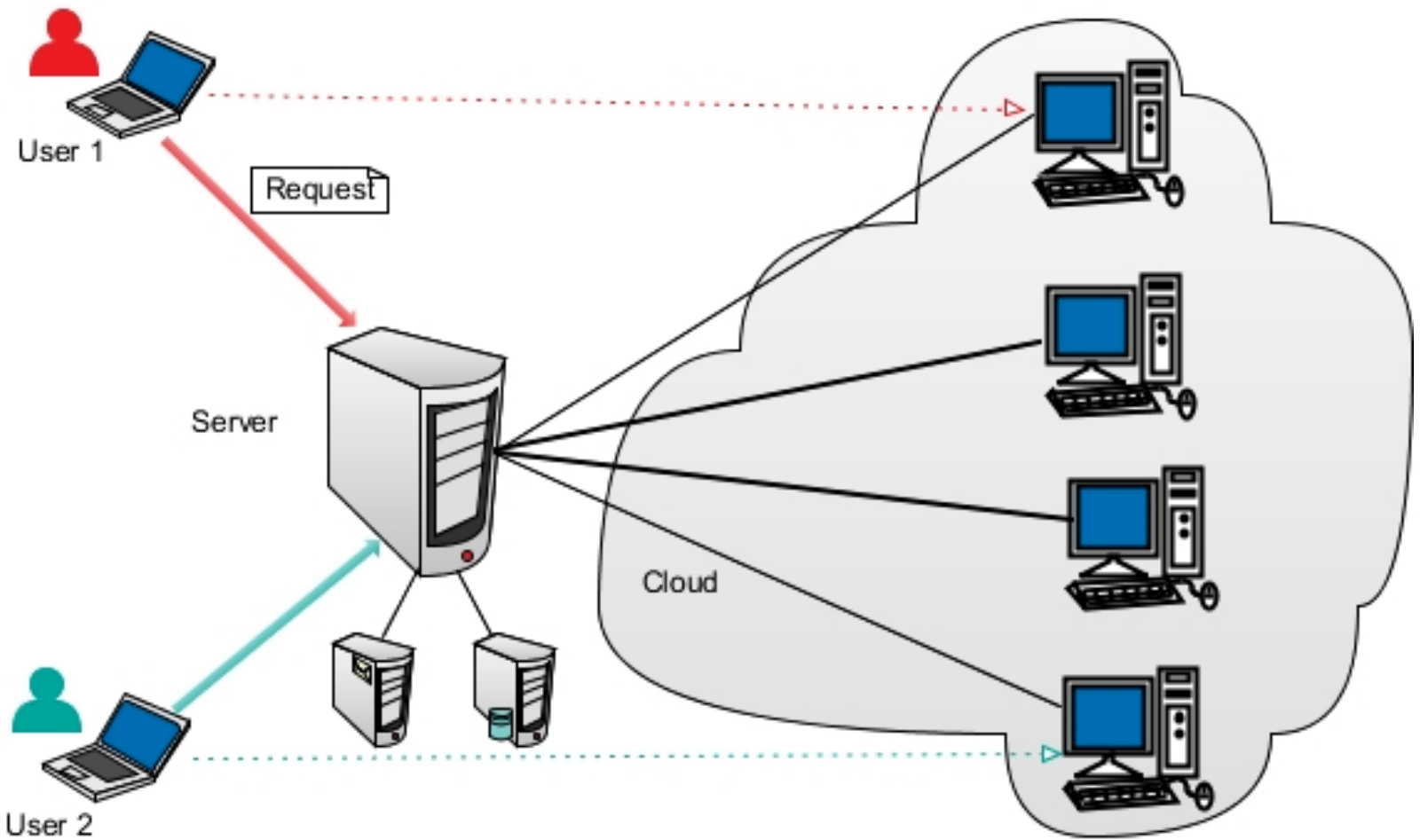

*Figure 6.1 Block diagram of the architecture*

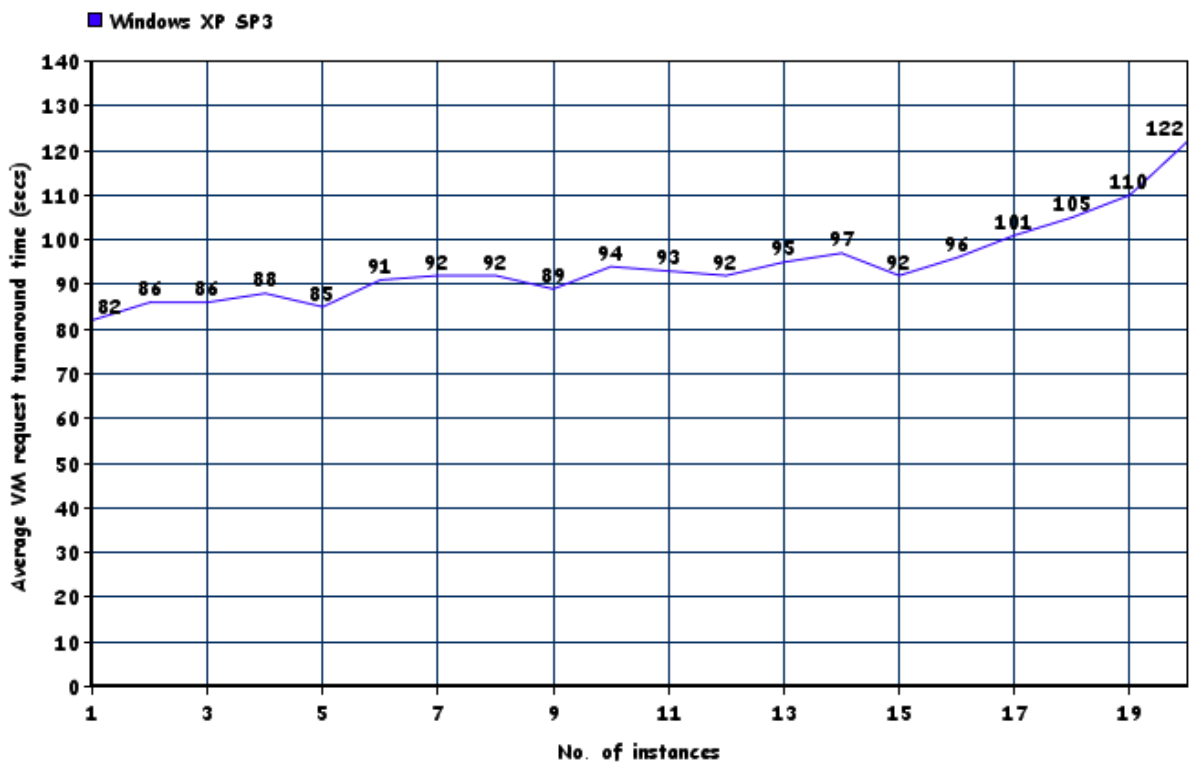

*Figure 2: Request turnaround time vs Load for Windows XP*

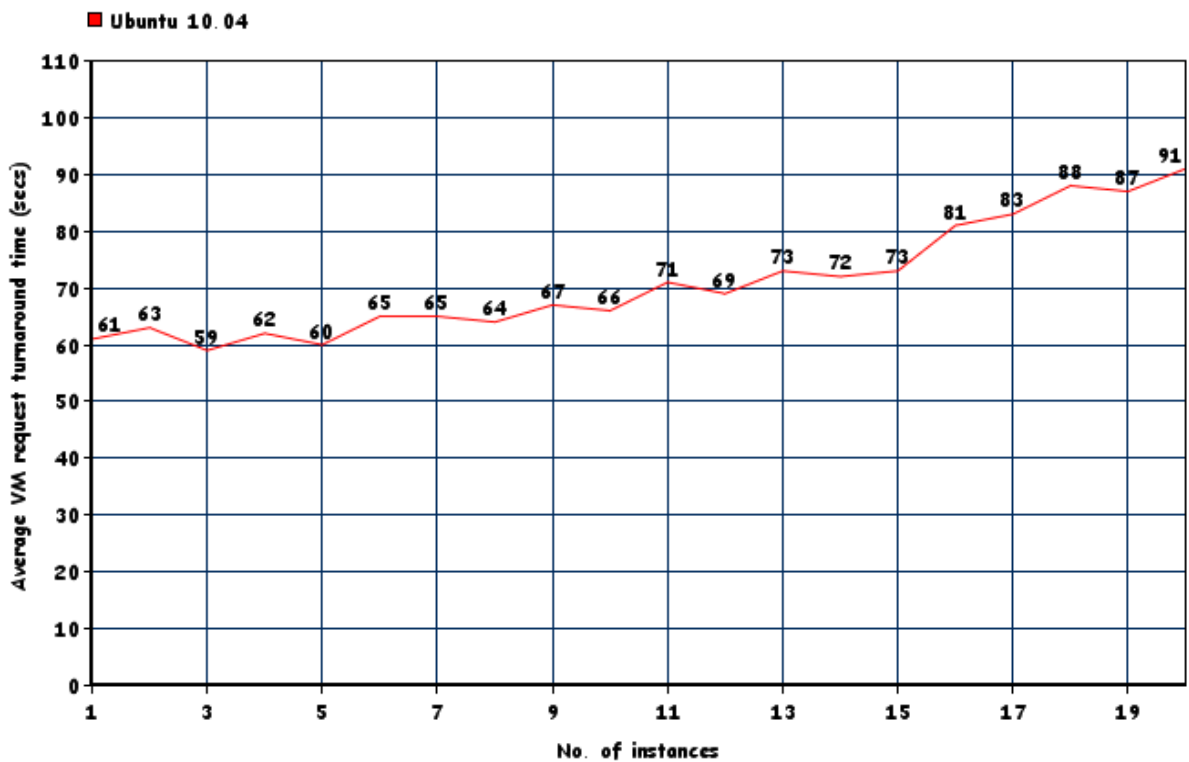

*Figure 2: Request turnaround time vs Load for Ubuntu 10.04*